\newcommand{\pp}{\ensuremath{\mathrm {p\kern-0.05em p}}}
\newcommand{\PbPb}{\ensuremath{\mbox{Pb--Pb}}}
\newcommand{\GeVc}{\ensuremath{\mathrm{GeV}\kern-0.05em/\kern-0.02em c}}
\newcommand{\sqrts}{\ensuremath{\sqrt{s_{\mathrm{NN}}}}}

\newcommand{\pT}{\ensuremath{p_{\mathrm{T}}}}

\newcommand{\pTtrack}{\ensuremath{p_{\mathrm{T}}^{\mathrm{track}}}}

\newcommand{\pTjet}{\ensuremath{p_{\mathrm{T}}^{\mathrm{jet}}}}

\newcommand{\Raa}{\ensuremath{R_{\mathrm{AA}}}}
\newcommand{\Ncoll}{\ensuremath{N_{\mathrm{coll}}}}

\newcommand{\Taa}{\ensuremath{\left<T_{\mathrm{AA}}\right>}}

\documentclass{PoS}

\usepackage{lineno}

\title{Inclusive jet measurements in pp and \PbPb{} collisions with ALICE}

\ShortTitle{Inclusive jet measurements in pp and \PbPb{} collisions with ALICE}

\author{James Mulligan\footnote{Speaker.} $\;$on behalf of the ALICE Collaboration\\
        Wright Laboratory, Department of Physics, Yale University, New Haven, CT 06520, USA\\
        E-mail: \email{james.mulligan@yale.edu}}

\abstract{
Measurements of jet yields in heavy-ion collisions can be used to constrain jet energy loss models, 
and in turn provide information about the physical properties of deconfined QCD matter.
ALICE reconstructs charged particle jets ({\it charged jets}) with high-precision tracking of charged particles down to $p_{\mathrm{T}}^{\mathrm{track}} = 150$ MeV/$c$,
and jets ({\it full jets}) with the addition of particle information from the electromagnetic calorimeter down to $E_{\mathrm{cluster}} = 300$ MeV.
By including low momentum jet constituents, ALICE is uniquely positioned at the LHC to measure 
jets down to low jet momentum, to determine the modification to the soft components of jets, and to measure medium recoil particles.
New inclusive full jet measurements in pp and Pb-Pb collisions at $\sqrt{s_{\mathrm{NN}}}=5.02$ TeV with ALICE will be shown,
over $R=0.2-0.4$ and extending to low jet $p_{\mathrm{T}}$. These will include the jet $R_{\mathrm{AA}}$ 
for different jet $R$, and will constitute the first such full jet measurements
at low transverse jet momentum at this collision energy. The results are compared to several theoretical predictions. }

\FullConference{International Conference on Hard and Electromagnetic Probes of High-Energy Nuclear Collisions\\
		30 September - 5 October 2018\\
		Aix-Les-Bains, Savoie, France}

\begin{document}



\paragraph{Introduction}

A deconfined state of Quantum Chromodynamics (QCD) is produced in ultra-relativistic heavy-ion collisions -- 
and the study of jet modification is one of the major avenues of the heavy-ion experimental program.
Jets traverse a significant pathlength of the medium, and the effect that the medium has on jets can be deduced by comparing jet properties in heavy-ion collisions to those in \pp{} collisions.
Previous measurements demonstrate suppression of the jet transverse momentum (\pT) spectrum in heavy-ion collisions, indicating that jets transfer 
energy to the hot QCD medium \cite{fullJet276, hjetPbPb, atlas502, jetRaa276CMS}.
However, the basic nature of this deconfined QCD state remains largely unknown.
Jets are sensitive to a wide range of momentum exchanges with the medium, 
and thereby can provide insight into the medium at a wide range of resolution scales.

We report measurements of inclusive jet \pT{} spectra in \pp{} and \PbPb{} collisions at $\sqrts=5.02$ TeV with the ALICE detector \cite{aliceDetector} at the Large Hadron Collider (LHC) \cite{LHCmachine}.
In \pp{}, we report the jet cross-section for resolution parameters $R=0.2,0.4$ over the range $20<\pTjet<140$ GeV/$c$.
In \PbPb{}, we report the $R=0.2,0.4$ jet spectrum for $40<\pTjet<140$ GeV/$c$ and $60<\pTjet<140$ GeV/$c$, respectively.
Jets are reconstructed at pseudo-rapidity $|\eta|<0.7-R$, and are required to contain at least one charged track with $\pTtrack>5-7$ GeV$/c$ (depending on the jet radius) in order to reject combinatorial jets. 
The jet spectra are fully corrected for detector and background effects.


\paragraph{Data analysis}

The reported \PbPb{} (\pp) data were recorded by the ALICE detector at the LHC in 2015 (2017) at $\sqrts=5.02$ TeV.
We utilize a sample of 4.5M (500M) 0-10\% \PbPb{} (\pp) accepted minimum bias events.
Jets are reconstructed with the FastJet 3.2.1 anti-$k_{\mathrm{T}}$ algorithm \cite{antikt} from the combination of 
charged particle tracks with $\pTtrack>150 \;\mathrm{MeV}/c$ and electromagnetic calorimeter (EMCal) clusters with $E_{\mathrm{cluster}}>300\; \mathrm{MeV}$.
We account for the fact that charged particles deposit energy in both the tracking system and the EMCal by extrapolating tracks to the EMCal and subtracting 
transverse momentum from the matched clusters. 

In \PbPb{}, we subtract the average background from each jet: $p_\mathrm{T,jet}^{\mathrm{reco}} = p_\mathrm{T,jet}^{\mathrm{raw}} - \rho A$, where
$\rho$ is the event-averaged background density in each event, and $A$ is the jet area \cite{fullJet276}.
However, $p_\mathrm{T,jet}^{\mathrm{reco}}$ fails to account for fluctuations in the underlying background and
a variety of detector effects, including tracking inefficiency, missing long-lived neutral particles, and material interactions.
We therefore deconvolute the reconstructed jet spectrum with a response matrix describing the correlation of $p_\mathrm{T,jet}^{\mathrm{reco}}$ to the true \pTjet{}, 
obtained by embedding a PYTHIA 8 Monash 2013 event with the GEANT3 ALICE detector simulation into \PbPb{} data. 
We then employ the SVD unfolding algorithm \cite{svd} using RooUnfold \cite{roounfold},
and correct the resulting spectrum for the kinematic efficiency and jet reconstruction efficiency.

We categorize two classes of systematic uncertainties: correlated uncertainties, 
which are positively correlated among all \pTjet{} bins, and shape uncertainties, which alter the shape of the final \pTjet{} spectrum.
The dominant correlated uncertainty is the uncertainty in the tracking efficiency, and the dominant shape uncertainty is the systematic uncertainty in the unfolding procedure.


\paragraph{Results}

\begin{figure}[!h]
\centering{}
\includegraphics[scale=0.37]{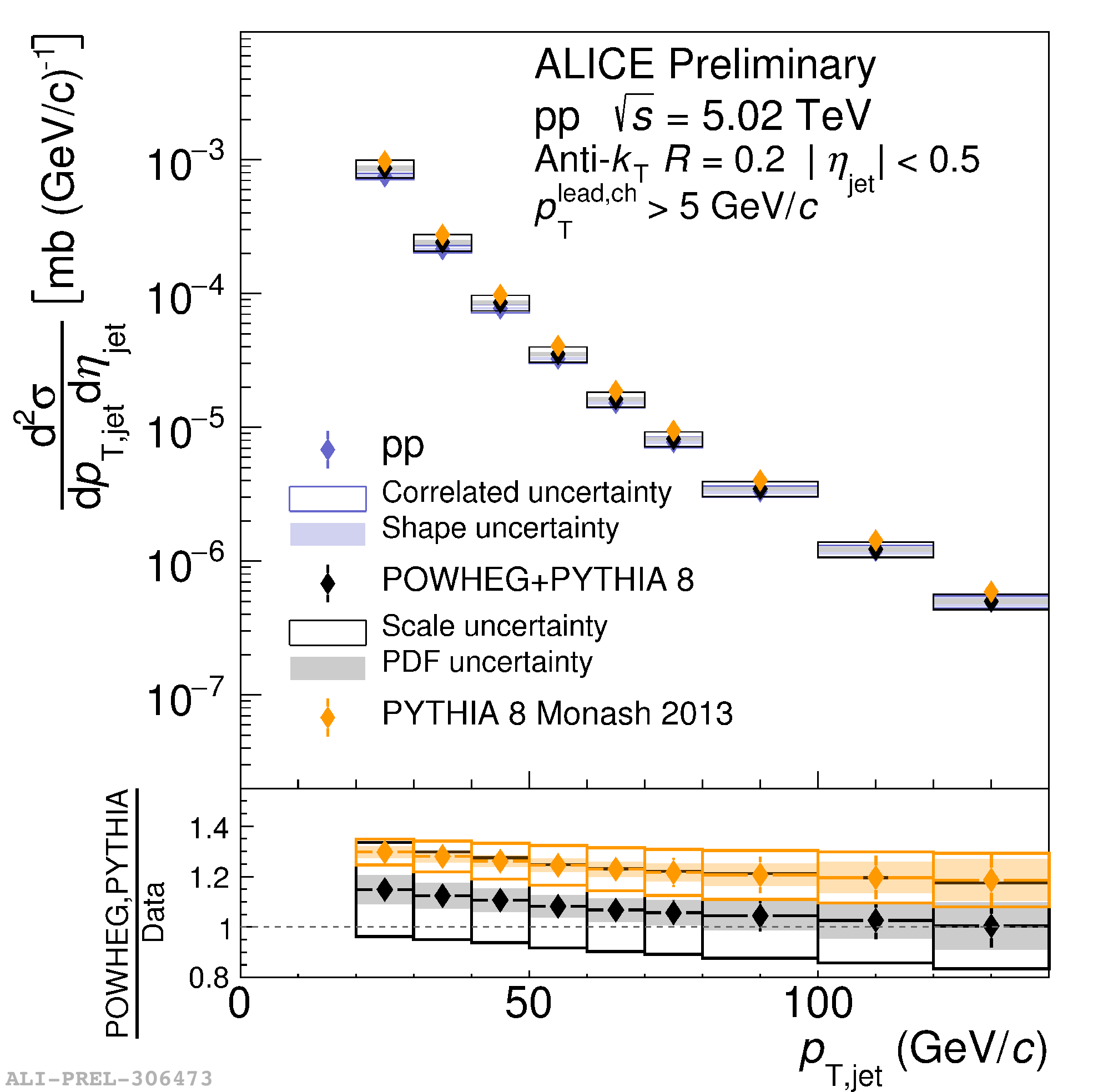}
\includegraphics[scale=0.37]{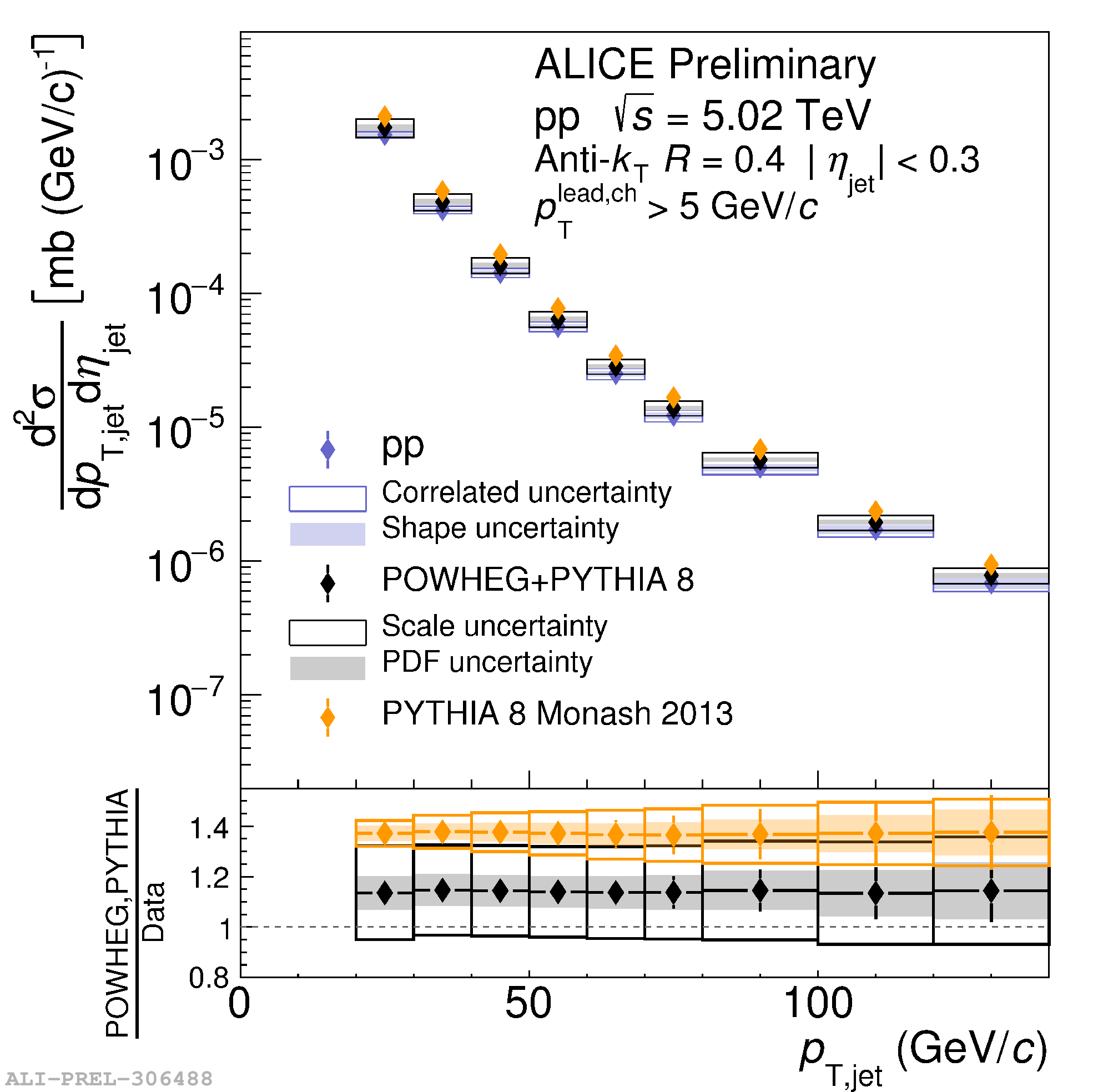}
\caption{Unfolded \pp{} full jet cross-section at $\sqrt{s}=5.02$ TeV for $R=0.2$ (left) and $R=0.4$ (right), along with PYTHIA 8 Monash 2013 and POWHEG+PYTHIA reference.
}
\label{fig:spectraPP}
\end{figure}

The \pp{} jet cross-sections are reported differentially in \pT{} and $\eta$ as: 
$\frac{d^{2}\sigma_{jet}}{d\pT d\eta} = \frac{1}{\mathcal{L}} \frac{d^{2}N}{d\pT d\eta},$
where we experimentally measure the yield $\frac{d^{2}N}{d\pT d\eta}$ and the integrated luminosity $\mathcal{L}$ \cite{ppXsec}.
The yield is corrected for the partial azimuthal acceptance of the EMCal and the vertex efficiency.
Figure \ref{fig:spectraPP} shows the unfolded \pp{} jet spectrum for $R=0.2$ and $R=0.4$ jets.
The jet cross-section predictions by PYTHIA 8 tune Monash 2013 are also plotted for comparison, 
as well as the NLO event generator POWHEG \cite{powheg1}, with PYTHIA 8 tune ATLAS-A14 for fragmentation.
The POWHEG predictions are consistent with the measured data, while PYTHIA 8 Monash 2013 alone is not.

\begin{figure}[!h]
\centering{}
\includegraphics[scale=0.37]{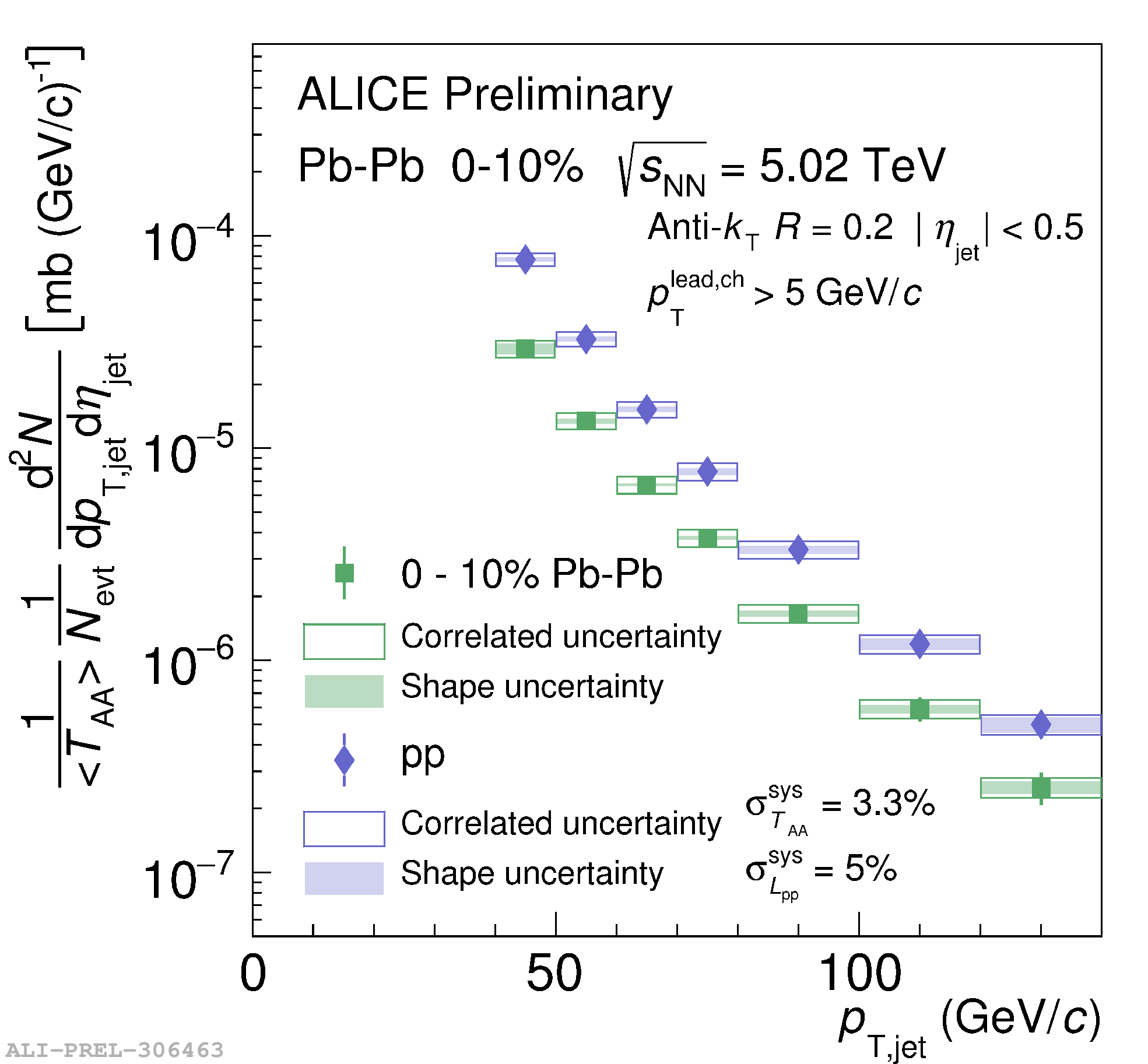}
\includegraphics[scale=0.37]{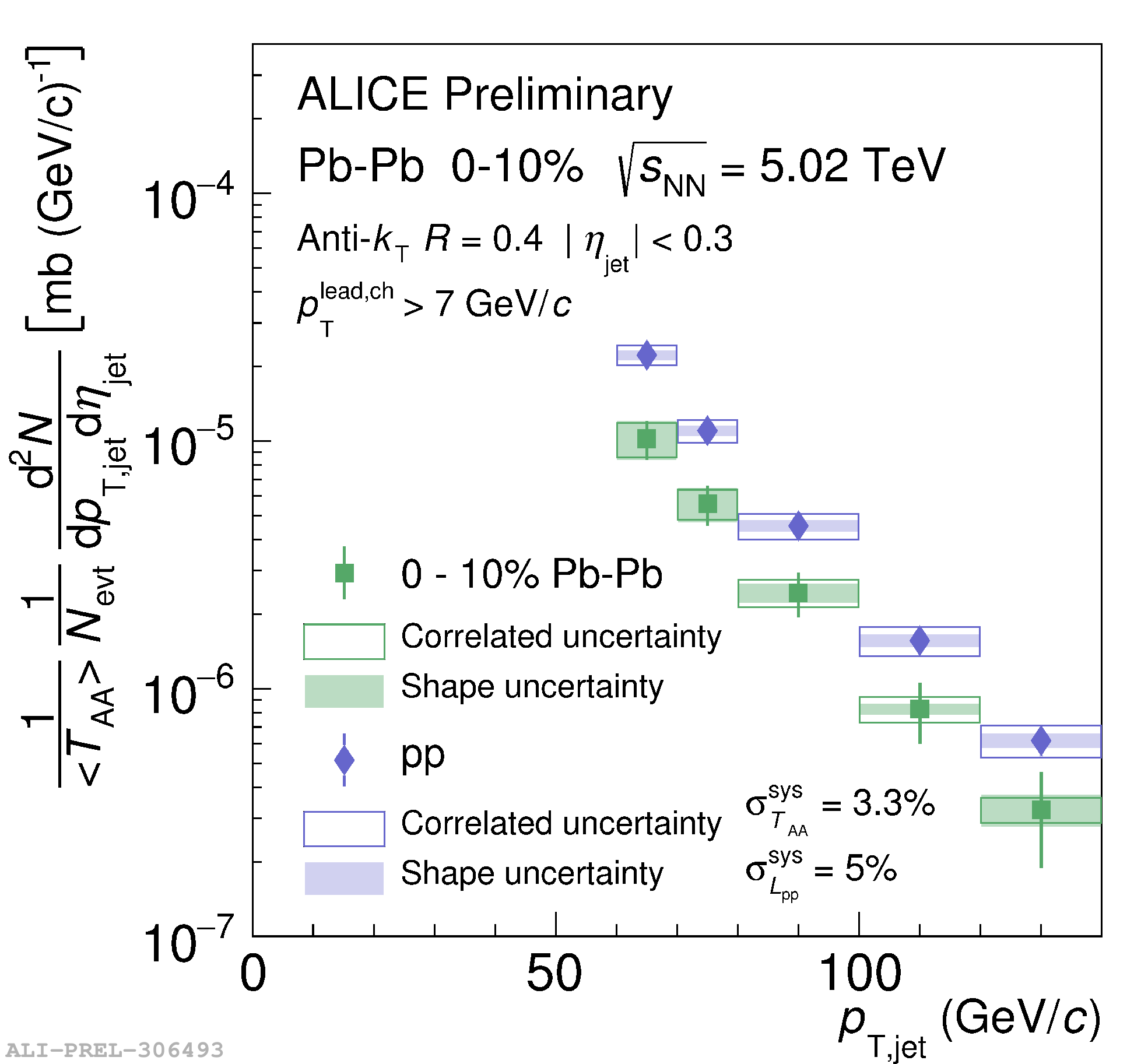}
\caption{Unfolded \pp{} and \PbPb{} full jet spectra at $\sqrts=5.02$ TeV for $R=0.2$ (left) and $R=0.4$ (right).}
\label{fig:spectra02}
\end{figure}

The \PbPb{} jet spectra are reported differentially in \pT{} and $\eta$ as: $\frac{1}{\Taa} \frac{1}{N_{event}} \frac{d^{2}N_{jet}^{AA}}{d\pT d\eta},$
where $\Taa \equiv \frac{\left<\Ncoll\right>}{\sigma_{inel}^{NN}}$ is the ratio of the number of binary nucleon-nucleon collisions 
to the inelastic nucleon-nucleon cross-section, computed in a Glauber model to be $\Taa = 23.4 \pm 0.78 \; \mathrm{(sys) \; mb^{-1}}$ for 0-10\% centrality.
Figure \ref{fig:spectra02} shows the unfolded \PbPb{} full jet spectra for $R=0.2$ and $R=0.4$. 
A leading track bias of 5 GeV/$c$ is required for the $R=0.2$ spectra, while a 7 GeV/$c$ bias is required for the $R=0.4$ spectra (both \pp{} and \PbPb{})
in order to suppress combinatorial jets in \PbPb.

\begin{figure}[!b]
\centering{}
\includegraphics[scale=0.37]{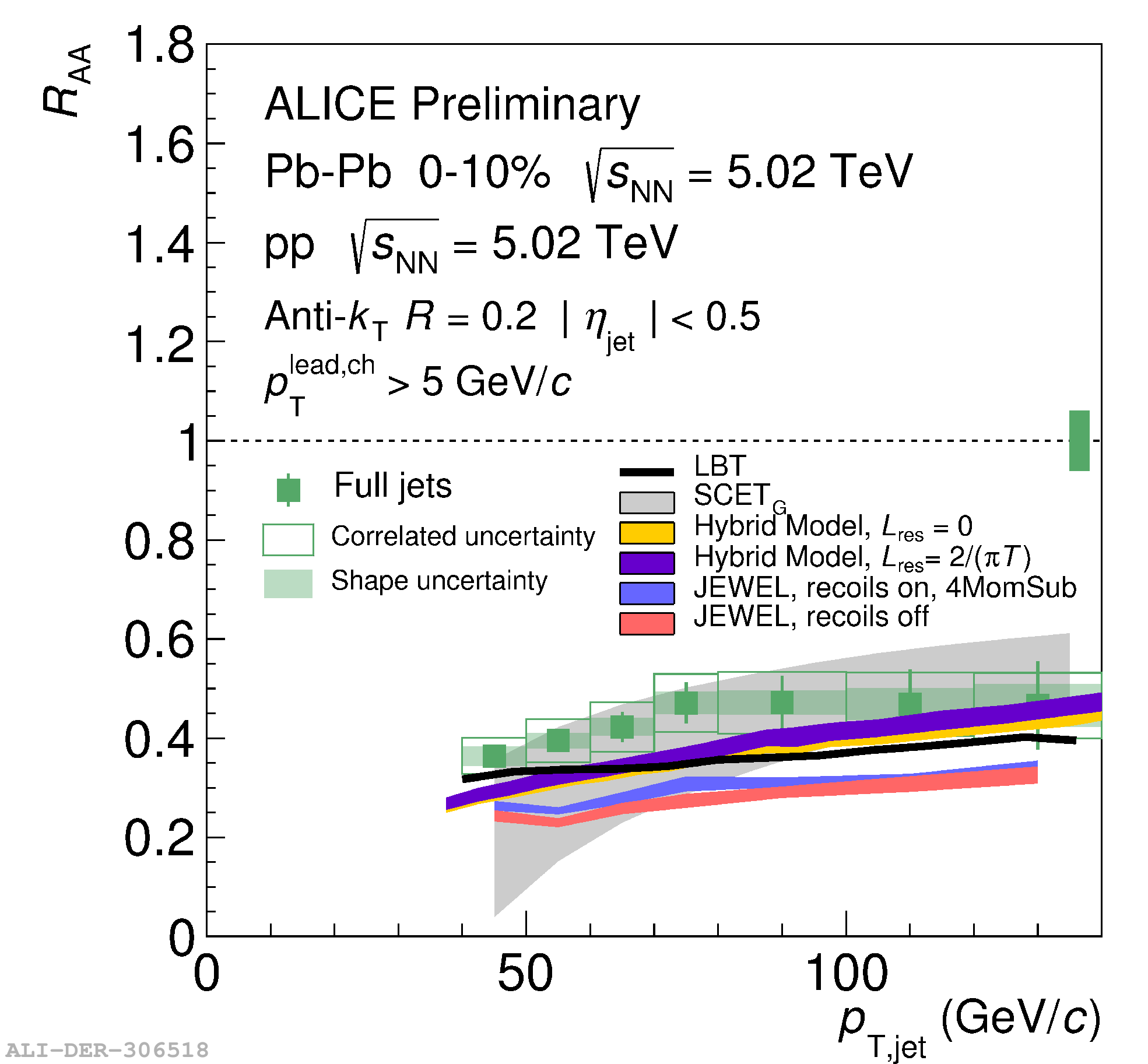}
\includegraphics[scale=0.37]{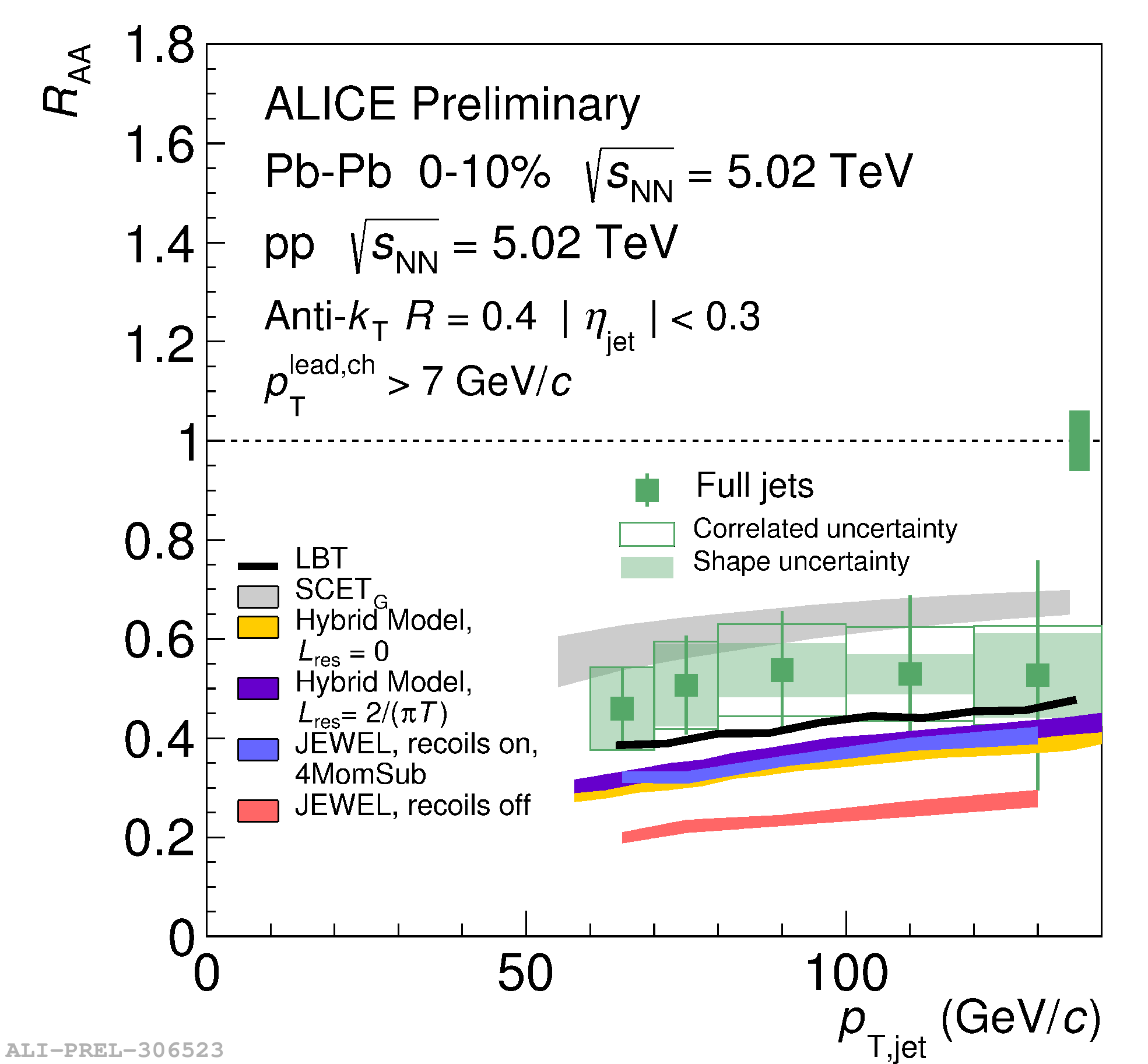}
\caption{Jet \Raa{} at $\sqrts=5.02$ TeV for $R=0.2$ (left) and $R=0.4$ (right) compared to LBT, SCET$_{G}$, Hybrid model, and JEWEL predictions. 
The combined \Taa{} uncertainty and \pp{} luminosity uncertainty of 6\% is shown as a band on the dashed line at $\Raa = 1$.}
\label{fig:Raa}
\end{figure}

The jet \Raa{} is reported as:
$\Raa = \left. { \frac{1}{\Taa} \frac{1}{N_{\mathrm{event}}} \frac{d^{2}N}{d\pT d\eta}}\right| _{\mathrm{AA}} \; \left/ \; {\frac{d^{2}\sigma}{d\pT d\eta} }\right| _{\pp},$
namely the ratio of the \PbPb{} and \pp{} spectra plotted in Fig. \ref{fig:spectra02}.
Figure \ref{fig:Raa} shows the unfolded full jet \Raa{} for $R=0.2$ and $R=0.4$ jets, which exhibit strong suppression.
There is visible \pT-dependence in the $R=0.2$ case, with stronger suppression at lower \pT. 
There is no significant $R$-dependence of the jet \Raa{} within the experimental uncertainties. 

We compare these results to four theoretical predictions: the Linear Boltzmann Transport (LBT) model \cite{LBT, LBTconspiracy}, 
Soft Collinear Effective Theory with Glauber gluons (SCET$_{G}$) \cite{SCET, vitev}, the Hybrid model \cite{HybridModel, dani}, and JEWEL \cite{Jewel, JewelVJet}. 
The predictions are all computed using the anti-$k_{\mathrm{T}}$ jet algorithm with $|\eta|<0.7-R$.
Leading track requirements are only applied by JEWEL (as in data) and the Hybrid model (with 5 GeV/$c$ for both radii).
All models exhibit strong suppression, and produce the same qualitative trend of \Raa{} as a function of \pT{}.
In the case $R=0.2$, we see that JEWEL under-predicts the jet \Raa{}, and appears to be inconsistent with the data regardless of whether medium recoils are included, while
for $R=0.4$ the ``recoils on" prediction is more consistent with the data.
The LBT model describes the data better, although it has slight tension with the data.
Note however that neither the JEWEL nor LBT predictions include systematic uncertainties.
The SCET$_{G}$ predictions are fully consistent with the data, although the $R=0.2$ prediction has large systematic uncertainties due to a lack of in-medium $\ln R$ re-summation in this calculation.
Additionally, the SCET$_{G}$ calculation did not include collisional energy loss, which the authors anticipate to increase the suppression for $R=0.4$.
The Hybrid model describes the trend of the data reasonably well, although like the LBT model, exhibits slight tension particularly in the $\pT < 100$ GeV/$c$ range.
It should be noted that JEWEL has no free parameters in the fit, and so it faces the strictest test of all the models presented.
While the experimental uncertainties are larger for $R=0.4$, the model predictions span a wider range of \Raa{} than in the case of $R=0.2$, 
which highlights the importance of measuring the $R$-dependence of the jet \Raa.

\clearpage
Most of the models describe the \Raa{} reasonably well, but a firm quantitative conclusion remains somewhat nebulous. 
The predictions typically use different strategies for each of the "non jet energy loss" pieces (initial state, expansion, hadronization, \pp{} spectrum),
and do not attempt to incorporate these differences in a systematic uncertainty, which makes a strict quantitative comparison to data difficult.
Moreover, the models fix their free parameters in different ways. 
This necessitates investigation of complementary jet observables and the need for global analyses, 
but it also highlights the need to standardize the ingredients of jet energy loss calculations.

\bibliographystyle{JHEP}
\bibliography{skeleton}

\end{document}